\documentclass{pasj00}
\twocolumn
\begin{document}
\SetRunningHead{Hashimoto et al.}
{Spatially Extended [P II]1.188 $\mu$m and [Fe II]1.257 $\mu$m in NGC 1068}
\Received{2010/10/05}
\Accepted{2010/11/29}
\Published{2011/02/25}

\title{Spatially Extended [P II]1.188 $\mu$m and [Fe II]1.257 $\mu$m Emission Lines 
in a Nearby Seyfert Galaxy NGC 1068 Observed with OAO/ISLE}



%
 \author{%
   Tetsuya \textsc{Hashimoto}\altaffilmark{1}
   Tohru \textsc{Nagao}\altaffilmark{2,3,4}
   Kenshi \textsc{Yanagisawa}\altaffilmark{5}
   Kenta \textsc{Matsuoka}\altaffilmark{2}
   and
   Nobuo \textsc{Araki}\altaffilmark{2}}
 \altaffiltext{1}{Department of Astronomy, Kyoto University, Kitashirakawa-Oiwake-cho, Sakyo-ku, Kyoto 606-8502}
 \email{tetsuya@kusastro.kyoto-u.ac.jp}
 \altaffiltext{2}{Graduate School of Science and Engineering, Ehime University, 2-5 Bunkyo-cho, Matsuyama 790-8577}
 \altaffiltext{3}{Research Center for Space and Cosmic Evolution, Ehime University, 2-5 Bunkyo-cho, Matsuyama 790-8577}
 \altaffiltext{4}{Optical and Infrared Astronomy Division, National Astronomical Observatory of Japan, Mitaka, Tokyo 181-8588}
 \altaffiltext{5}{Okayama Astrophysical Observatory, National Astronomical Observatory of Japan, Kamogata, Okayama 719-0232}


\maketitle

\begin{abstract}
We present $J$-band long-slit spectroscopic observation of NGC 1068 
classified as a Seyfert 2 galaxy. $J$-band observations with OAO/ISLE 
provide clear detection of spatially extended [Fe II]1.257$\mu$m 
and [P II]1.188$\mu$m lines. We found that 
[Fe II]1.257$\mu$m/[P II]1.188$\mu$m increases with distance from a 
central continuum peak. Observed line ratios around the nucleus (continuum peak) 
are consistent with a typical value expected from photoionization models, 
while the ratios at 3$\arcsec - 4\arcsec$ ($210-280$ pc) east and west of the nucleus 
are slightly higher than this. In the off nucleus region of NGC 1068 we also found 
a possible association between [Fe II]1.257$\mu$m/[P II]1.188$\mu$m and 
the radio continuum. This suggests a mild contribution of shock ionization 
induced by a radio jet outside nucleus while photoionization by the central energy 
source is dominant near the nucleus. 
\end{abstract}

\section{Introduction}
\label{introduction}

The narrow line regions (NLRs), which extend to several hundred or 
kilo parsec scale around the galaxy center, are the exclusive structure 
of active galactic nuclei (AGN) where the spatially resolved observations 
are possible. Therefore NLRs are often investigated as an important tool 
to study the ionization state of the interstellar medium (ISM) and/or chemical 
evolution in galactic scale (e.g., \cite{2006A&A...447..863N}). 
Although it is widely accepted that the NLR is photoionized by ionizing 
photons radiated from a central engine, the possibility of shock ionization 
induced by a jet in off nucleus regions cannot be excluded since ionization 
photons decrease with distance from a nucleus (e.g., \cite{2007ApJ...666..794F}). 
Thus, how much the shock contributes to the ionization of NLRs is very important 
to our understanding of AGN structure and in examining the utility of NLRs as a tool 
to investigate galactic-scale phenomena. 

Furthermore, the shock ionization of 
NLRs is getting a lot more attention. Recent dramatic progress of theoretical 
simulations and observational studies of galaxy formation and evolution allow 
a quantitative comparison between both sides. In this context, a serious problem 
has arisen, i.e., theoretical simulations predict too many massive galaxies due 
to long-duration star formations in contrast to early-time quenching of star 
formation in observed massive galaxies (e.g., \cite{2006MNRAS.365...11C,2006MNRAS.370..645B}). 
This problem cannot be solved even if a negative feedback effect on star 
formation activity caused by supernovae is involved and so AGN feedback effect 
is considered as a potential candidate of a solution: a massive galaxy likely has 
a supermassive black hole at its galaxy center, and inflow of ISM to a supermassive 
black hole invokes its AGN activity which releases vast gravitational potential energy 
to ISM resulting in suppression of star formation activity 
(e.g., \cite{2005ApJ...635L..13S,2007MNRAS.380..877S}). However, how the 
AGN activity transmits its energy to ISM remains a mystery. One possible 
physical mechanism of the AGN feedback is shock ionization of ISM, i.e., 
the AGN activity inputs its energy to ISM through a shock heating induced by 
a jet. 

In previous studies of NLRs, line-ratio diagnostics to distinguish between shock 
ionization and photoionization have been examined. Optical diagnostics, however, 
can hardly discriminate between the two mechanisms, because optical NLR spectra 
predicted by photoionization and shock ionization models are very similar to each 
other \citep{1995ApJ...455..468D,1996ApJS..102..161D,2008ApJS..178...20A}. 
The near infrared line ratio of [Fe II]1.257$\mu$m/[P II]1.188$\mu$m is 
one of the most powerful indicators to discriminate photoionization and shock 
ionization. Both lines have similar critical densities and excitation temperatures, 
i.e., this line ratio is roughly proportional to the ratio of gas-phase abundance 
of iron and phosphorous. In contrast, iron is a well known refractory species and 
is strongly depleted in dust grains, whereas phosphorous is a non-refractory species. 
Photoionization alone (including H II regions and NLRs excited by ionizing photons 
from young stars and AGN central sources, respectively) is relatively incapable of 
destroying the tough iron based grains, while these are easily sputtered by shocks. 
The [Fe II]1.257$\mu$m/[P II]1.188$\mu$m ratio, therefore, is high 
($\gtrsim 20$) in fast shock-excited regions and low ($\lesssim 2$) in normal 
photoionized regions 
\citep{2001A&A...369L...5O}. The actual ionization state of NLRs would be determined 
by the combination of photoionization and shock ionization, and so the observed line 
ratios are expected to change with the locations in NLRs ranging from 
[Fe II]1.257$\mu$m/[P II]1.188$\mu$m $\sim 2$ to $\sim 20$. 

NGC 1068 is one of the nearest AGNs, which has a compact radio jet around the
nucleus and spatially extended radio lobe \citep{1983ApJ...275....8W}. This
radio structure well coincides with a morphology of NLR 
\citep{1997ApJ...487..560C}. Therefore, NGC 1068 is an ideal object with which to 
research the spatial distribution of [Fe II]1.257$\mu$m/[P II]1.188$\mu$m. 
In this paper, we adopt 14.4 Mpc as a distance to NGC 1068 and 
1$\arcsec$ corresponds to 70 pc. 

\section{Theoretical [Fe II]1.257$\mu$m/[P II]1.188$\mu$m ratio}
Before describing the details of observation and data reduction, we summarize 
the theoretical background of [Fe II]1.257$\mu$m and [P II]1.188$\mu$m 
emission lines.

The emission line intensity is proportional to the product of the density 
of the ion responsible for the emission line process ($n_{i}$) and the electron 
density ($n_{e}$), multiplied by a function $f$ 
giving the rate of the process. Thus the intensity ratio of 
two emission lines radiated from ion 1 and 2 is written as
\begin{equation}
\frac{I(\lambda _{1})}{I(\lambda _{2})} = 
\frac{n_{i,1}f_{1}}{n_{i,2}f_{2}},
\end{equation}
assuming the same spatial distribution of both ions \citep{1989agna.book.....O}. 
The function $f$ involves the rate of emission line photons in the radiative 
transition from the excited level to the ground level written as 
\begin{equation}
\frac{n_{1}A_{10}}{n_{0}} = A_{10}\frac{q_{01}(T)}{q_{10}(T)} 
\left [1+\frac{A_{10}}{n_{e}q_{10}(T)} \right]^{-1}.
\end{equation}
Here $A_{10}$ is the radiative transition probability, and $q_{01}$ and $q_{10}$ 
are collisional excitation and deexcitation rate, which include collisional strength 
($\Omega$). Thus, $f$ can be calculated if $A$ and $\Omega$ are known. 
\cite{2001A&A...369L...5O} derived 
\begin{equation}
\frac{n({\rm Fe})}{n({\rm P})} \lesssim 
\frac{n({\rm Fe}^{+})}{n({\rm P}^{+})} \sim 2\cdot 
\frac{I([{\rm Fe\ II}]1.257\mu{\rm m})}{I([{\rm P\ II}]1.188\mu{\rm m})}, 
\label{abundance}
\end{equation}
using the collision strengths and transition probabilities of 
[Fe II]1.257$\mu$m and [P II]1.188$\mu$m 
\citep{1970RSPSA.318..531K,1995A&A...293..953Z,1982MNRAS.199.1025M,1988A&A...193..327N}
. For solar abundance ratio of $n({\rm Fe})/n({\rm P}) \sim 100$ and typical 
depletion factor of Fe ($\sim 0.01$) and P ($\sim 1.0$ because of refractory 
species), equation (\ref{abundance}) gives 
[Fe II]1.257$\mu$m/[P II]1.188$\mu$m close to unity, although the 
depletion factor of iron differs from object to object. 
Actually, [Fe II]1.257$\mu$m/[P II]1.188$\mu$m
is $\lesssim 2$ in normal photoionized region, 
e.g., $\sim 2$ in Orion Bar \citep{2000A&A...364..301W}. This is also true 
for NLR ionized by AGN radiation because even ionizing photons from AGN central 
source can hardly destroy the tough iron based grains in NLR. 

However if shocks exist the grains are easily destroyed and gas-phase 
iron increases. As a result shock ionized gas represents high 
[Fe II]1.257$\mu$m/[P II]1.188$\mu$m ratio. If we assume that 
iron based grains are completely sputtered by shocks, the ratio becomes 
$\sim 50$ for solar abundance. This is similar to that measured in 
supernova remnants (e.g., $\gtrsim 20$ for LMC-N63A and LMC-N49 
reported by \cite{2001A&A...369L...5O}). 

The actual observed line ratio in NLR of AGN is expected to be between 
$\sim 2$ and $\sim 20$ since the ionization state would be determined 
by the combination of photoionization and shock ionization if these exist 
as mentioned in section \ref{introduction}. 

\section{Observations and data reduction}
\label{Observation}
\begin{figure}
  \begin{center}
    \FigureFile(80mm,80mm){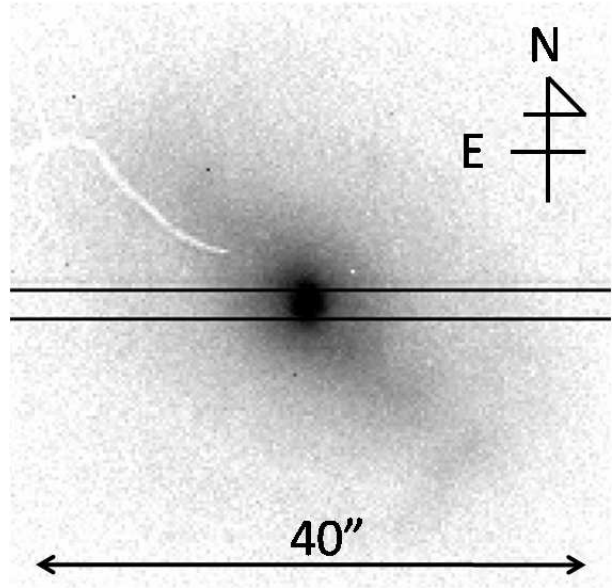}
  \end{center}
  \caption{
$J$-band image of NGC 1068 obtained with OAO/ISLE in our observation. 
The long-slit position 
(P.A. = 90$^{\circ}$) is shown by two solid lines. 
}\label{figure1}
\end{figure}

Long-slit spectroscopy was carried out from November 8 to 12 2009 with ISLE 
\citep{2006SPIE.6269E.118Y,2008SPIE.7014E.106Y}, 
which is a near-infrared imager and spectrograph for the Cassegrain focus of the 
1.88 m telescope at Okayama Astrophysical Observatory (OAO). 
The camera used for the spectroscopic observations has a projected scale of 
0$\arcsec$.25/pixel. The spectrum of NGC 1068 was obtained with a slit of 
2$\arcsec$.0 (= 8 pixels) width and $J$-band grating which yields a 
$1.11 - 1.32\mu$m spectrum with a dispersion of 0.166 $\mu$m/pix.  
The spectral resolution is $\sim$ 1300 measured from an OH emission line at 
the central wavelength. 
The slit was oriented to E-W (i.e., position angle = 90$^{\circ}$) and 
centered on the $J$-band continuum peak of NGC 1068 (Fig. \ref{figure1}). 
We note that the position angle is fixed to 90$^{\circ}$ in ISLE spectroscopic mode. 
Therefore, the slit was not placed along the major axis of NLR and the 
direction of the radio structure, at a position angle of $\sim$ 30 $^{\circ}$ 
\citep{2006AJ....132..620D,2010ApJ...708..419C}. 
It lies outside of the nominal bicone of NLR and away from the axis of radio emission. 
Unfortunately other slit positions north and south of the nucleus could not 
be completed due to bad weather conditions. 
The acquisition consisted of 
a series of two 2-minute exposures with the object set at different positions 
along the slit followed by dome flats and calibration lamp of Argon and Xenon. 
The seeing size was 1$\arcsec$.0 - 2$\arcsec$.0. Since the weather conditions 
were not good throughout the observation, we excluded 
poor data from a total 6-hour exposure on source, resulting in the effective 
exposure time of 4.4 hours on source. 
The standard data reduction 
was performed for all selected spectra, i.e., dark frame subtraction, flat fielding, 
wavelength calibration, and sky subtraction using IRAF software. 
To correct the atmospheric spectral response and the instrumental efficiency, 
spectra of NGC 1068 were divided by spectra of some A-type rationing stars 
(HIP5310, HIP10795, HIP14077, and HIP22774) 
with the same airmass. In this reduction, the black body and Pa$\beta$ absorption 
features of rationing stars were removed by spectral fitting with a black body 
function and Voigt profile, respectively. The assumed effective temperatures of 
rationing stars are 8270, 7500, 8200, and 9230 K for HIP5310 (A3V), 
HIP10795 (A7V), HIP14077 (A5V), and HIP22774 (A1V), respectively.

\section{Results and discussion}
\label{Results}
\begin{figure}
  \begin{center}
    \FigureFile(160mm,80mm){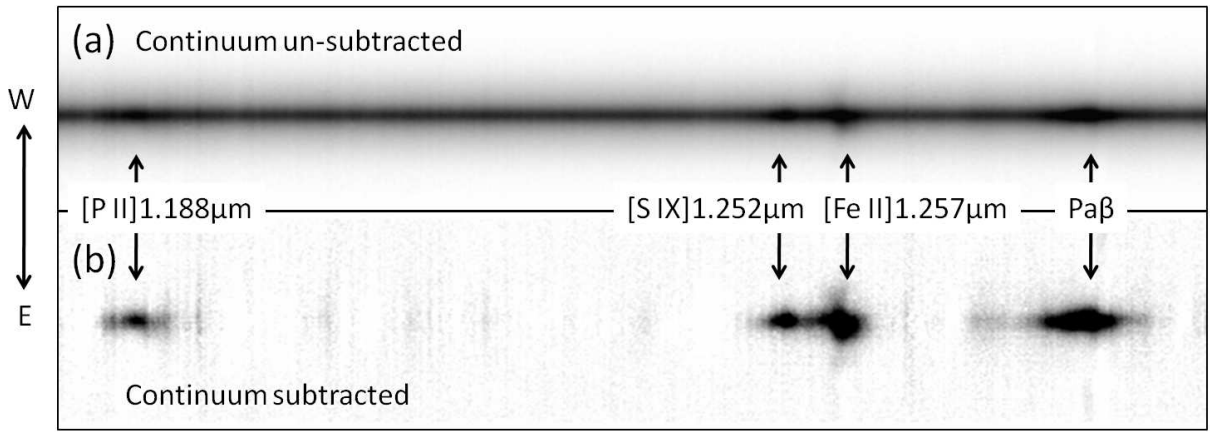}
  \end{center}
  \caption{
2-D spectra in $J$ band extracted from central $\pm$15$\arcsec$ region (a) and 
continuum-subtracted spectrum (b). 
}\label{figure2}
\end{figure}

The obtained 2-D spectra of NGC 1068 are displayed in Fig. \ref{figure2}. 
We detected [Fe II]1.257$\mu$m 
and [P II]1.188$\mu$m lines as well as Pa$\beta$ and [S IX]1.252$\mu$m. 
The spatial extent of [Fe II]1.257$\mu$m and [P II]1.188$\mu$m are 
$\sim 14\arcsec$ and $\sim 7\arcsec$. These values 
are clearly larger than the typical seeing size of $1\arcsec .0-2\arcsec .0$. 
Thus, we concluded that spatially extended [Fe II]1.257$\mu$m and 
[P II]1.188$\mu$m were successfully detected. 

\begin{figure}
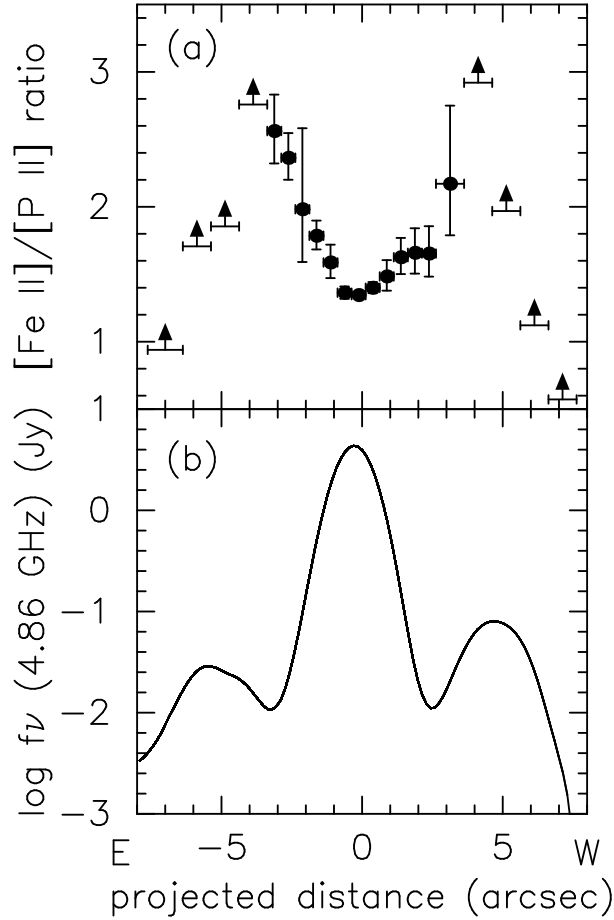

  \begin{center}
    \FigureFile(80mm,80mm){figure-3.eps}
  \end{center}
  \caption{
[Fe II]1.257$\mu$m/[P II]1.188$\mu$m line ratio (top) and VLA 4.86 GHz flux density 
(bottom) as a function of distance from a continuum peak. Arrows in the top figure are 
lower limits calculated from 3 $\sigma$ noise level around undetected [P II]1.188$\mu$m. 
}\label{figure3}
\end{figure}

The line fluxes were measured from spectral fitting analysis with IRAF $specfit$ 
task \citep{1994asp...61...437}, assuming single gauss and underlying linear 
functions for each emission line. We summarized relative line fluxes 
normalized by [P II]1.188$\mu$m in Table \ref{table}, extracted from 
central 2$\arcsec$.0 region and east and west neighbor regions. 
The detailed spatial distribution of 
[Fe II]1.257$\mu$m/[P II]1.188$\mu$m line ratios is shown in Fig. 
\ref{figure3} (a). 

\cite{2001A&A...369L...5O} reported that the line ratio in the 
central $\sim$ 2$\arcsec$ region of NGC 1068 is about 1.5. 
Since this value corresponds with the photoionization scheme as mentioned above, 
they concluded that in the central region most iron is locked into grains and 
shock excitation is not the primary origin of [Fe II] line emission. This explanation 
is relatively straightforward because there would be a large number of ionizing 
photons near the nucleus, that is enough to dominate the ionization of surrounding 
gases. Our measurement of [Fe II]1.257$\mu$m/[P II]1.188$\mu$m $\sim$ 1.3 in 
the central 2$\arcsec$ region is consistent with this value.  

However, this argument may not be valid at off nucleus regions. We found that 
[Fe II]1.257$\mu$m/[P II]1.188$\mu$m increases with 
distance from a central continuum peak. While observed line ratios around the 
nucleus are consistent with a prediction by photoionization models, 
the ratios at 3$\arcsec$ $-$ 4$\arcsec$ east and west of the nucleus 
($\sim$ 560 pc) are slightly higher than the typical value of 
[Fe II]1.257$\mu$m/[P II]1.188$\mu$m in the photoionized region. 
Although there are only two research efforts devoted to the spatial distribution of 
[Fe II]1.257$\mu$m/[P II]1.188$\mu$m ratios of NLRs 
(NGC 4151 by \cite{2009MNRAS.394.1148S} 
and Mrk 1066 by \cite{2010MNRAS.404..166R}), similar results 
were reported in both cases. \cite{2009MNRAS.394.1148S} 
found that [Fe II]1.257$\mu$m/[P II]1.188$\mu$m ratios are higher at 
$\sim$ 130 pc away from the nucleus of NGC 4151 ($\sim 6$) than that in its 
nucleus ($\sim 2$). They also pointed out a possible spatial correlation 
between [Fe II]1.257$\mu$m/[P II]1.188$\mu$m and radio continuum structure. 
They suggested that shocks induced by a radio jet release the Fe locked in grains 
and produce an enhancement of the [Fe II] emission at off nucleus regions. 
Similarly \cite{2010MNRAS.404..166R} found that Mrk 1066 presents 
[Fe II]1.257$\mu$m/[P II]1.188$\mu$m $\sim 3$ at most locations within 
$\sim 470$ pc from the nucleus, but in some regions close to the borders of the 
radio continuum structure this ratio reaches values up to 9.5. They concluded 
that shocks seem to play a more important role in these regions.

Fig. \ref{figure3} (b) shows VLA 4.86 GHz flux density as a function of distance 
from the nucleus of NGC 1068 extracted from a same slit aperture as our 
near-infrared observation with OAO/ISLE. The radio data was obtained from 
NRAO Science Data Archive\footnotemark[1].
\footnotetext[1]{https://archive.nrao.edu/archive/e2earchivex.jsp}
At off nucleus region of NGC 1068 we find 
a possible association between [Fe II]1.257$\mu$m/[P II]1.188$\mu$m and 
the radio continuum like NGC 4151 and Mrk 1066. The higher ratios at off nucleus of 
NGC 1068 is likely attributed to a mild contribution of shock ionization to ionized 
gases. This may indicate that the interaction between 
the jet and ISM forms an expanding cocoon which induces the shock waves propagating 
perpendicularly in the direction of the jet axis (e.g., \cite{1974MNRAS.166..513S}), 
while photoionization by central engine is dominant near the nucleus. 

\section{Conclusion}
The line ratio [Fe II]1.257$\mu$m/[P II]1.188$\mu$m in the near-infrared 
wavelength range is a useful tool with which to examine the dust destruction by shocks. 
We investigated spatial distribution of this ratio in NLR of nearby Seyfert galaxy 
NGC 1068 with OAO/ISLE. 
[Fe II]1.257$\mu$m/[P II]1.188$\mu$m near the nucleus is close to unity 
consistent with a previous observation and with a ratio in a normal photoionized 
region. This indicates that photoionization by ionizing photons radiating from a central 
engine is dominant near the nucleus. We found that the ratio increases with the 
distance from the nucleus, and is slightly higher at 3$\arcsec$ $-$ 4$\arcsec$ 
east and west of the nucleus than ratios typical of a photoionized region. 
We also found a possible spatial association between 
[Fe II]1.257$\mu$m/[P II]1.188$\mu$m and radio continuum around $\sim 560$ pc 
from the nucleus. These findings suggest a higher contribution of shock ionization 
induced by a radio jet at off nucleus. Except for NGC 1068, recently the spatial 
correlation between [Fe II]1.257$\mu$m/[P II]1.188$\mu$m and radio continuum 
over the several hundred parsec scale has been reported for NGC 4151 and Mrk 1066. 
Applying this kind of research to a number of other AGNs is the clue to revealing 
ongoing AGN feedback phenomena. \\

We would like to thank Nozomu Kawakatu for his meaningful comments on 
interpretation of observed data. This work was supported by the 
Publications Committee of the National Astronomical Observatory of Japan (NAOJ) and 
the Grant-in-Aid for the Global COE Program \lq\lq The Next Generation of Physics, 
Spun from Universality and Emergence\rq\rq from the Ministry of Education, Culture, 
Sports, Science and Technology (MEXT) of Japan. 
T.N. acknowledges financial supports through the Research 
Promotion Award of Ehime University and the Kurata Memorial 
Hitachi Science and Technology Foundation. 
K.M. acknowledges financial support from the Japan 
Society for the Promotion of Science (JSPS) through 
the JSPS Research Fellowships for Young Scientists. 
\onecolumn

\onecolumn

\begin{table}
  \caption{Relative line fluxes normalized by [P II]1.188$\mu$m}\label{table}
  \begin{center}
    \begin{tabular}{lclclc|c|}
\hline
Line ID&East 3$\arcsec.0$&Central 2$\arcsec$.0&West 3$\arcsec$.0\\
\hline
${\rm [P\ II]}$1.188$\mu$m&1.0$\pm$0.29&1.0$\pm$0.02&1.0$\pm$0.08\\ 
${\rm [S\ IX]}$1.252$\mu$m&1.05$\pm$0.02&1.01$\pm$0.02&0.73$\pm$0.07\\
${\rm [Fe\ II]}$1.257$\mu$m&1.79$\pm$0.02&1.33$\pm$0.05&1.63$\pm$0.07\\
Pa$\beta$&2.35$\pm$0.04&2.89$\pm$0.05&2.55$\pm$0.08\\
\hline
    \end{tabular}\\
Spectra were extracted from central 2$\arcsec$.0 region and east and west neighbor 
3$\arcsec$.0 regions.
  \end{center}
\end{table}


\bigskip
\newpage



\end{document}